\documentclass[aps,prl,twocolumn,showpacs]{revtex4}
\usepackage{graphicx}
\usepackage{epsfig}
\usepackage{times}
\usepackage{color}

\renewcommand{\vec}[1]{\mbox{\boldmath$\mathrm{#1}$}}

\begin{document}

\title{Helical multiferroics for electric field controlled quantum  information processing}
\author{M. Azimi$^1$, L. Chotorlishvili$^1$, S. K. Mishra$^1$, S. Greschner$^2$, T. Vekua$^2$,
and J. Berakdar$^1$} \affiliation{
$^1$Institut f\"ur Physik, Martin-Luther-Universit\"at Halle-Wittenberg,06120 Halle, Germany\\
$^2$Institut f\"ur Theoretische Physik, Leibniz Universit\"at Hannover, 30167 Hannover, Germany}

\begin{abstract}
Magnetoelectric coupling in helical multiferroics allows to steer spin order with electric fields.
Here we show theoretically  that in a helical multiferroic chain  quantum information processing as well as quantum phases are highly sensitive to electric ($E$) field. Applying $E$-field, the quantum state transfer fidelity can be increased and made  directionally dependent.
 We also show that  $E$ field transforms the spin-density-wave/nematic or multipolar phases of frustrated ferromagnetic spin$-\frac{1}{2}$ chain in chiral phase with a strong magnetoelectric coupling. We find sharp reorganization of the entanglement spectrum as well as a large enhancement of fidelity susceptibility at Ising quantum phase transition from nematic to chiral states driven by electric field.
 These findings point to a new tool for quantum information with low power consumption.
\end{abstract}
\date{\today }

\maketitle
\emph{Introduction}.-
Multiferroics (MF) are materials that show simultaneously multiple  spontaneous ferroic ordering  \cite{schmidt}.
Intrinsic coupling between the order parameters, e.g.  ferromagnetism (FM), ferroelectricity (FE), and/or ferroelasticity (for an overview we refer to \cite{ME-review_prl,Single-ME_prl,Composite-ME_prl,Zavaliche-11_prl,MeKl11_prl,Wang,Ramesh,Garcia,Bibes,6,picozzi,Dawber,Valencia,rev1,rev2,rev3,rev4,rev5,rev6,rev7}), allows
for multifunctionality of devices with  qualitatively new  conceptions
\cite{ME-memory_prl,MTJ1_prl,MTJ2_prl,prl13,APL_jia}. Particulary  advantageous  is the high sensitivity
of some MF compounds to external fields \cite{field1,field2,field3,field4,field5}.  This allows to steer, for instance
 magnetic order with moderate  electric   fields opening thus the door
 for magnetoelectric spintronics  and spin-based information processing
  with  ultra low power consumption and  dissipation \cite{ME-memory_prl,MTJ1_prl,MTJ2_prl,APL_jia}.
 These prospects are fueled by  advances in synthesis and nano fabrication  that render feasible
   versatile MF nano and quantum structures with enhanced  multiferroic coupling
\cite{ME-review_prl,Single-ME_prl,Composite-ME_prl,Zavaliche-11_prl,MeKl11_prl,fiebig13}.
  From a fundamental point of view MF
are also fascinating as  their properties  often emerge from  an interplay of
competing exchange and electronic correlation, crystal symmetry, and coupled  spin-charge dynamics.
 For example,
 the  perovskite multiferroics  RMnO$_3$ with R = Tb, Dy, Gd and Eu$_{1-x}$Y$_x$  exhibit
an incommensurate chiral spin order \cite{RMnO3} coupled to a finite  FE polarization. The underlying physics is
 governed by  competing exchange   and  Dzyaloshiskii-Moriya (DM) \cite{Dz59} interactions:
The spin-orbital coupling 
 associated with $d(p)$-orbitals of  magnetic(oxygen) ions triggers the
 FE  polarization  \cite{KNB,Jia} $\mathbf{P} \propto \hat{e}_{ij} \times (\mathbf{S}_i \times \mathbf{S}_j)$;
   $\hat{e}_{ij}$ is a unit vector  connecting   the sites $i$ and $j$  at which the
    effective spins $\mathbf{S}_{i/j} $ reside (e.g., along [110] direction for  TbMnO$_3$).
    $\mathbf{P}$  is thus linked with the spin order chirality $\kappa=(\mathbf{S}_i \times \mathbf{S}_j)$, offering thus  a tool for
    electrical control of $\kappa$ because, as shown experimentally \cite{Yama}, $\mathbf{P}$
     can be steered with
  an external electric field $E$ (with $|E|\sim$ 1 kV/cm).
   Indeed, effects of magnetoelectric coupling (ME) are  evident in the dynamical response to moderate $E$ \cite{ME-RMnO3,EuYMnO3-1,EuYMnO3-2,d-GL,GdMnO3,EPLjia}, i.e. phenomena rooted in ME can be driven, and possibly controlled
    by moderate external fields. Noteworthy, the chiral behaviour of  TbMnO$_3$ persists with  miniaturization   down to 6nm \cite{fiebig13}. Furthermore, the feasibility was demonstrated of  multiferroic spin $1/2$ chain of  LiCu$_2$O$_2$ \cite{park07}  and field-switchable LiCuVO$_4$ \cite{schrett08}.

 These facts combined with the robust topological nature of the intrinsic chirality are the key elements
 of the present proposal to utilize chiral MF for $E$-field controlled, spin-based quantum information processing.
Starting from an established model \cite{KNB} for  chiral MF with the aim to inspect electrically driven quantum information processing  and quantum phases in a multiferroic chain,  we
find that electric field $E\sim 10^{3}$kV/cm increases strongly the quantum state transfer fidelity  making it direction dependent.
 The system can be steered electrically between spin-density-wave, nematic, multipolar and chiral phases. We find an E- field modifies drastically
 the entanglement spectrum and an enhances the fidelity susceptibility at Ising quantum phase transition from nematic to chiral states.

\emph{Theoretical framework.-}
We employ an effective model with frustrated spin interaction for the description of a one-dimensional MF chain along the $x$ axis \cite{KNB}.
 The chain is subjected to an electric ($E$, applied along the $y$ axis) and a magnetic ($B$ along the $z$ axis) fields.  The Hamiltonian reads
\begin{eqnarray}
\label{eq.2}
\hat{H}=J_1\displaystyle\sum_{i=1}\vec{S}_i.\vec{S}_{i+1}
+J_2\displaystyle\sum_{i=1}\vec{S}_i.\vec{S}_{i+2}-B\displaystyle\sum_{i=1}S_{i}^{z}-\vec{E}\cdot\widehat {\vec{P}}\, .
\end{eqnarray}
The exchange interaction constant between nearest neighbor spins is chosen FM $J_1<0$, while next-nearest interaction is antiferromagnetic $J_2>0$.
Below, we use units in which $J_2=1$ (typical values, e.g., for LiCu$_2$O$_2$ are  $J_1\approx -11 \pm 3$ meV; and $J_2\approx 7 \pm 1$ meV \cite{mfprl21,mfprl22,mfprl23,mfprl24}).
Eq. (\ref{eq.2}) is an effective Hamiltonian based on the conditions that
  $E(B)$ fields are weak such that their direct coupling to electronic orbital motion is negligible.
  The classical $E$ field couples (with a strength $g_{ME}$) to the induced
  polarization, i.e.,  $\vec{E}\cdot \widehat{\vec{P}}=Eg_{ME} \sum_{i}(\vec{S}_{i}\times
\vec{S}_{i+1})^z$. While $\vec{S}_{i}$ will be treated  fully quantum mechanically, displacements
 will not be quantized \cite{EPLjia}.  $\kappa=\langle \kappa_i\rangle =\langle (\vec{S}_{i}\times
\vec{S}_{i+1})^z\rangle $ is known as $z$ component of vector chirality (VC), which  for brevity we call chirality.\\
The frustrated  $J_1-J_2$ spin$-\frac{1}{2}$ chain was studied extensively both theoretically \cite{chubukov91a,kolezhuk05,hm06a,vekua07,kecke07,hikihara08,sudan09} ( exhibiting  its rich ground state phase diagram hosting multipolar and chiral phases) and experimentally \cite{enderle05,enderle10,drechsler07}.
 However,  neither the control of quantum information processing via  external driving fields nor the effect of electric field on the ground state properties have been addressed yet. The present study is a contribution to fill these gaps.

We note that the electric field coupling  term resembles a Dzyaloshinskii-Morija (DM) anisotropy, with a coupling constant $d=g_{ME}E$.  Experiments indicate the presence of a small DM anisotropy in MF cuprates made of frustrated spin  chains \cite{enderle05,enderle10};  previous theories  considered it negligible, however.
 Here we show that even a tiny DM anisotropy modifies considerably  the spin $1/2$ chain characteristics. In particular, nematic spin-density-wave (SDW) state of magnon as well as multipolar phases transform into a chiral Luttinger liquid with non-zero spin current in the ground state.\\
First we focus analytically on a minimal system of four spins for different strengths of magnetic and electric (driving) fields for
 establishing an efficient protocol to field-control the entanglement.
We also inspect quantum state transfer fidelity (QSTF) through  MF chain and its $E$-field dependence.

For strong $B$-fields, i.e. $B$ is larger than  $|J_{1}|$,  $1$, and $d=g_{ME}E$, the ground state  is fully polarized, namely  $\vert F \rangle=| \uparrow \uparrow \uparrow  \uparrow\rangle$.
The corresponding energy is  ${\cal E}_F=J_1+1-2B$. The pair entanglement between any two arbitrary spins and the chirality vanish.
Decreasing the magnetic field so that  $B_0<B<d+J_1+2$, where $B_0= \sqrt{(J_1-4)^2+8d^2}/2+(J_1-2d)/2 $,
the ground state  is  $\vert\psi_{1}\rangle=\frac{i}{2}\vert  \downarrow \uparrow \uparrow  \uparrow  \rangle+\frac{-1}{2}\vert  \uparrow \downarrow \uparrow  \uparrow   \rangle
+\frac{-i}{2}\vert  \uparrow \uparrow \downarrow   \uparrow  \rangle+\frac{1}{2}\vert  \uparrow \uparrow \uparrow  \downarrow  \rangle$
with the corresponding energy ${\cal E}_{1}=-1-B-d$.
The chirality jumps to  $\kappa=\langle\psi_{1}\vert\displaystyle \kappa_i \vert\psi_{1}\rangle=1$. We
 observe a finite entanglement, as quantified by the pair concurrence between spins on $n$ and $m$ sites \cite{Amico}
$C_{nm}=max(0,\sqrt{R_1}-\sqrt{R_2}-\sqrt{R_3}-\sqrt{R_4})$, where  $R_n$  are the eigenvalues of the matrix $R=\rho_{nm}^{R}(\sigma_{1}^{y}\bigotimes\sigma_{2}^{y})(\rho_{nm}^{R})^{*}(\sigma_{1}^{y}\bigotimes\sigma_{2}^{y})$,
and $\rho_{nm}^{R}$ is the reduced density matrix of the  four spins system  obtained from
the density matrix  $\hat{\rho}$ after tracing over two  spins.
One can contrast the amount of the entanglement
stored in the pair correlations, quantified by the so-called two-tangle $\tau_2$,
 with the multi-spin entanglement of the whole spin
chain, encapsulated in  the  one-tangle, $\tau_1=4{\rm det }\rho_1$ \cite{Amico} ($\rho_1$ is the single spin reduced density matrix).
Two-tangle is calculated as $\tau_2=\sum_{m}^{4}C_{nm}^{2}$.
%
 For the state $\vert\psi_{1}\rangle$ we find the ratio $\tau=\frac{\tau_2}{\tau_1}=1$, thus half of the entanglement generated by decreasing the magnetic field (or increasing the electric field) in
$\vert\psi_{1}\rangle$ is stored in the collective multi-spin correlations and half in the pair correlations.
It is instructive  to study the effect of  $ E $ and  $B$ fields on  quantum-transfer fidelity, QSTF, \cite{Bose} between  different states,
\begin{eqnarray}\label{eq.8}
&&F(E,B,t)=\frac{\vert f_{j,s}(E,B,t)\vert\cos\gamma}{3}
+\frac{\vert f_{j,s}(E,B,t)\vert^2}{6}+\frac{1}{2},\nonumber
\\
&&\gamma=arg\{f_{j,s}(E,B,t)\}.
\end{eqnarray}
 $f_{j,s}(E,B,t)=\langle j\vert \exp(-i\hat{H}t)\vert s\rangle $ is the transition amplitude between the states $\vert j\rangle$ and $\vert s\rangle$.

Time dependencies of QSTF obtained analytically  between the initial state $ \vert1\rangle=\vert  \downarrow  \uparrow \uparrow \uparrow   \rangle $ and final states $\vert 2\rangle=   \vert \uparrow \downarrow \uparrow \uparrow  \uparrow  \rangle$ and $\vert 3\rangle=   \vert \uparrow \uparrow \downarrow \uparrow    \rangle $  are depicted  in the Fig.1.  The results evidence that
$E$-field increases QSTF, particularly from $ \vert1\rangle $ to $ \vert3\rangle $.
By inspecting (\ref{eq.8}) we infer that the oscillating behavior of $F$ in Fig.1.   is related to  the interference
effect between different quantum states ${\cal E}_{n}\big(E\big)/\hbar$. Note that electric field $E$ enters in the energy
levels through the DM coupling leading to a shift of state energies and the transition strength. For the explicit expression
of Fidelity see supporting materials.

 For confirmation we performed numerical calculations for  systems with a large number of spins (not shown) and  observed similar behavior of  QSTF on $ E $. We note that  $E$-field breaks the parity symmetry of the MF spin chain. Hence, when $E$-field is present, clockwise and anticlockwise QSTF between the states $\vert j \rangle \longrightarrow \vert s\rangle $ and $\vert s\rangle \longrightarrow \vert j\rangle$ differ considerably (cf. Fig.1, which
 might be used for information transfer control via magnetic chirality  \cite{Menzel}.
\begin{figure}[!h]
\includegraphics*[width=0.26\textwidth]{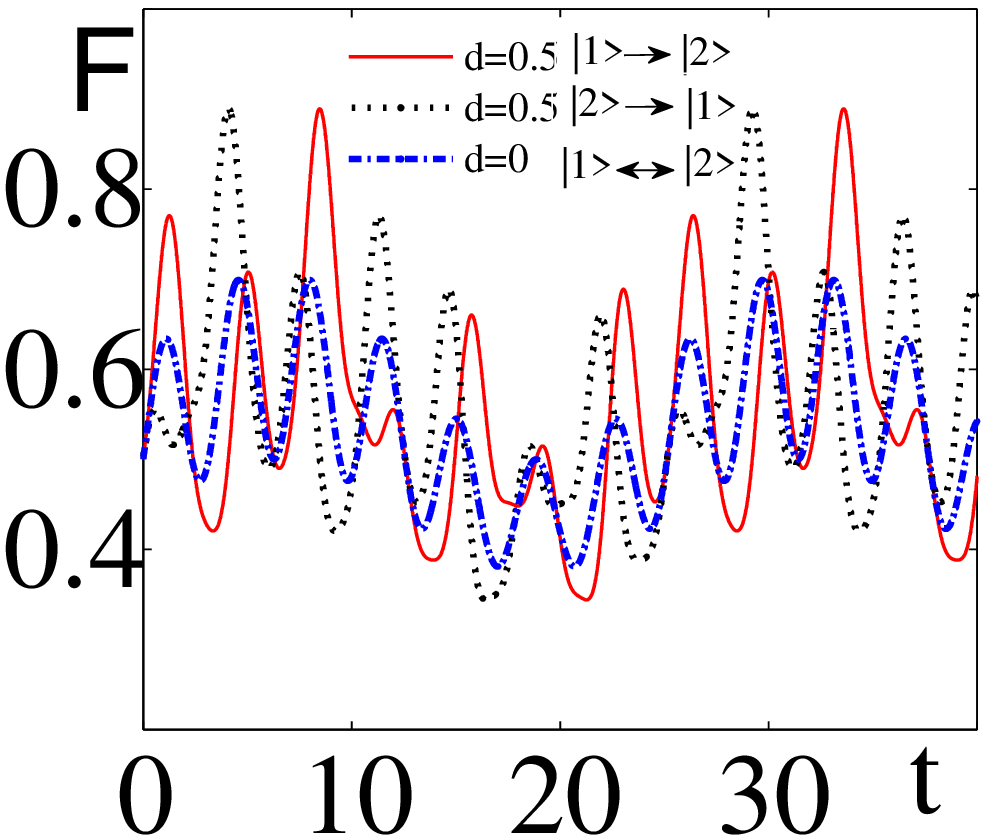}\includegraphics*[width=0.22\textwidth]{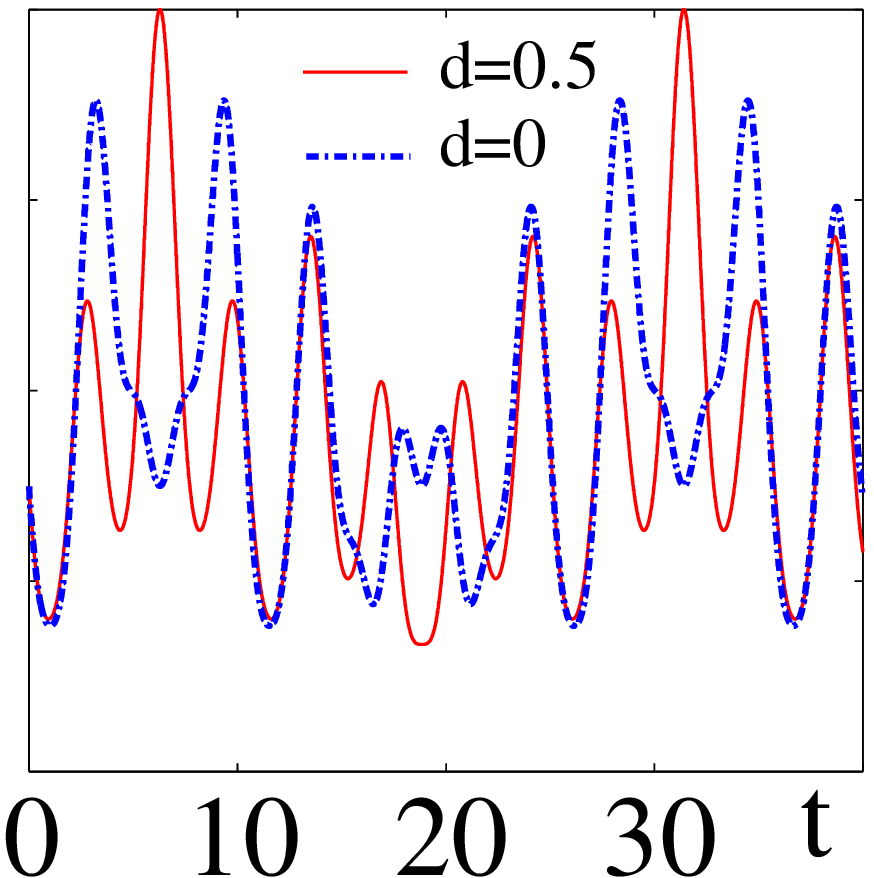}
\caption{(Color online) Time and $E$-field dependence of the QSTF in a four spin chain, as quantified by $F$ (cf.  eq.(\ref{eq.8})).
$d=E\, g_{ME}$.
  QSTF are depicted  on the left panel for the states:
  $ \vert 1\rangle  \to  \vert 2\rangle  $,
  $ \vert 2\rangle  \to  \vert 1\rangle$. Right panel shows
   QSTF  for $ \vert 1\rangle \to \vert 3\rangle =  \vert 3\rangle \to \vert 1\rangle $. We set dimensionless units
  $-J_1=J_2=1,B=1/4$. Time is measured in $\hbar/J_{2}$. In material parameters, e.g.  for  LiCu$_2$O$_2$  (cf. Ref. \cite{park07}),
and $\hbar/J_{2}=0.1[ps]$. $d=0.5$ corresponds to $E=10^{3}[kV/cm]$ assuming in a cell of size  $a_{FE}\approx 10[nm]$  a polarization of $P=P_{0}a_{FE}^{3}$
with $P_{0}=5\cdot 10^{-6}[C/m^{2}]$  (which is within the range measured in Ref.\cite{field4}.  As we choose  $S_{N+1}=S_{1}$, for $N=4$ the transition $ \vert 1\rangle \to \vert 3\rangle$  shows  no directional dependence for the fidelity.
  }
\label{Fig.1}
\end{figure}
Further decreasing the magnetic field below $B_0$, the ground state becomes
\begin{eqnarray*}
\vert\psi_2\rangle=\beta\big(\vert  \uparrow \uparrow \downarrow \downarrow   \rangle
-i\lambda\vert \downarrow \uparrow \downarrow \uparrow \rangle
-\vert  \uparrow \downarrow \downarrow \uparrow \rangle-\vert \downarrow \uparrow \uparrow \downarrow\rangle \\
+i\lambda\vert \uparrow \downarrow \uparrow \downarrow \rangle+\vert  \downarrow \downarrow \uparrow \uparrow \rangle\big),
\end{eqnarray*}

$\lambda =({J_1/4-1
+\sqrt{1+J_1^2/16-J_1/2+d^2/2}})/{\frac{d}{2}}$ and $\beta=1/\sqrt{4+2\lambda^2}$.
 In this case for chirality we have $\kappa=\langle\psi_{2}\vert \kappa_i\vert\psi_{2}\rangle=8\lambda\beta^{2}.$
and we plot its electric field dependence in Fig. 2 (a).
The ratio between one-tangle $\tau_1$ and two-tangle $\tau_2$
in the ground state $\vert\psi_{2}\rangle$ reads
$\tau=\frac{\tau_2}{\tau_1}=(\frac{2-\lambda^2}{2+\lambda^2})^2<1$, for $0<d\leq\frac{2-J_1/2}{7}$.
Therefore, in this case  the entanglement generated by the electric field
is  stored basically in many spin correlations rather then in two spin correlations.
\begin{figure}[h]\label{Fig.1}
\includegraphics*[width=0.185\textwidth]{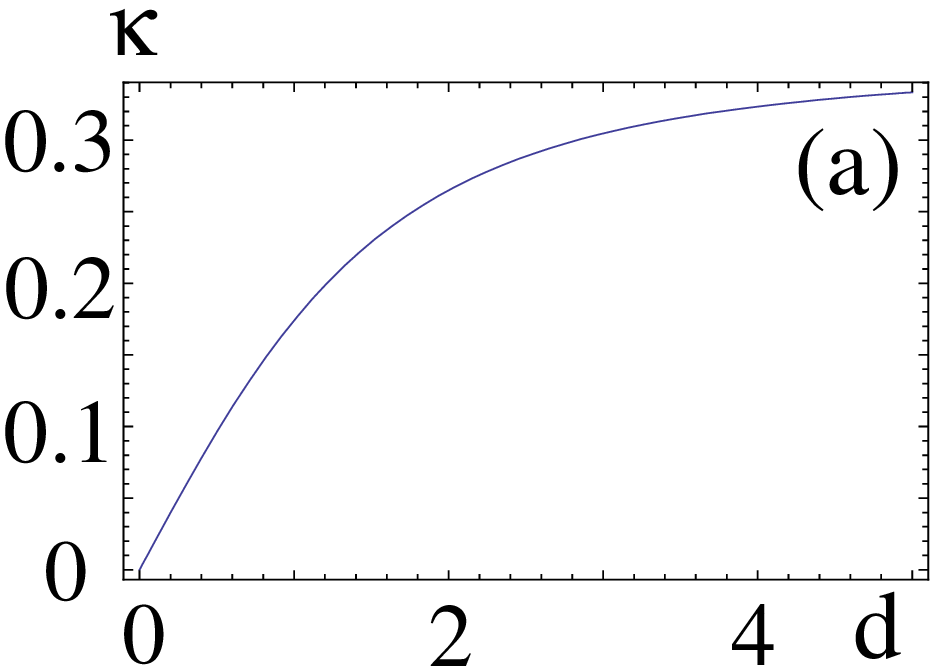}\includegraphics*[width=0.177\textwidth]{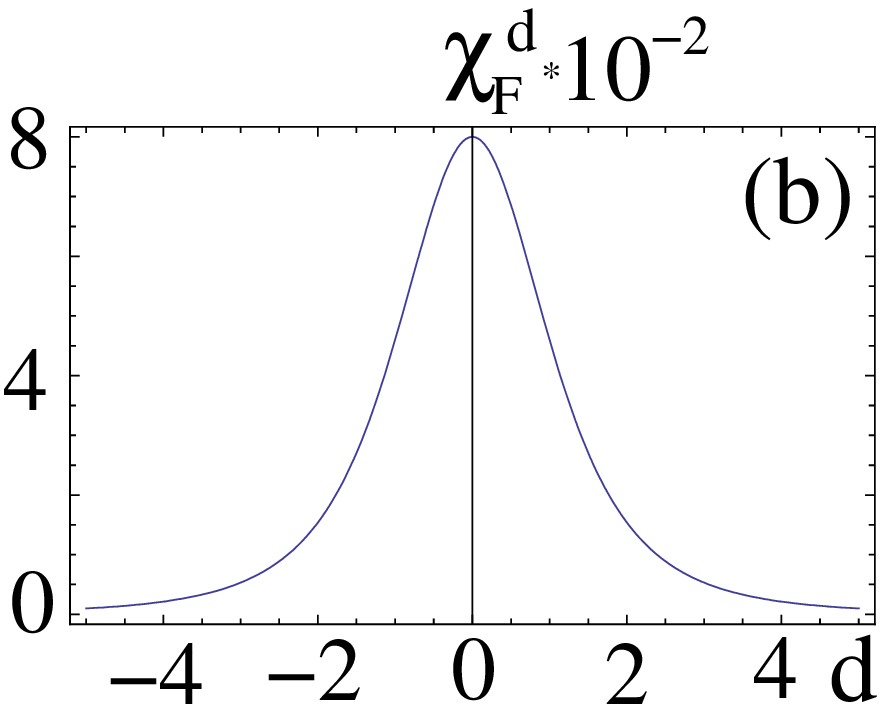}
\caption{a) Electric field dependence of chirality for the following values of the parameters $-J_1=J_2=1,B=1/4$. We see that electric field generates chirality. Qualitatively similar dependence holds even in thermodynamic limit. Electric field control of the magnetic chirality in the ferroaxial MF system $RbFe(MoO_4)_2$ was addressed in Ref.[\onlinecite{Hearmon}].  b) Electric field fidelity susceptibility.
As we see, due to the transition to the chiral phase, even a weak  electric field leads to a substantial reduction of the FS.}
\end{figure}

Response sensitivity  with changing the driving field amplitude is quantified by the  fidelity susceptibility (FS) \cite{Zanardi2}.
 FS with respect to magnetic field vanishes as the magnetization is conserved in our model.
FS with  E-field changes is finite. E.g., for $\vert\psi_{2}\rangle$ state we obtain: $\chi_{F}^{d}=\big(\alpha\beta/d\big)^2$
and depict it in Fig. 2 (b). As we see even small amplitude of the electric field leads to the substantial reduction of the FS.
Physical reason of the observed effect is transition to the chiral phase. We will study FS for long chains later, especially its behavior near the nematic to chiral quantum phase transition (QPT).

Hence depending on the driving fields, quantum information characteristics
such as many particle entanglement and QSTF differ considerably.
For macroscopic number of sites driving fields lead to different quantum phases and QPTs in frustrated FM chain.
 For MF chain we can expect thus a similar behavior that can   possibly be controlled by $E$ field. Hence, we study below $E$-field steered quantum phases and their transitions
  in a macroscopic MF chain.
 We focus on the thermodynamic limit. Before addressing the many-body physics it is instructive to start with the two-magnon problem: For $d=0$ and weak $J_1<0$  a bound state of two magnons forms below the scattering continuum. The bound state branch has a minimum for the total momentum $K=\pi$, for   antiferromagnetic $J_2$ disfavors two-magnons  occupying sites of the same parity.
We solved analytically the two-magnon problem for $d\neq 0$ (for  $L \to \infty$). The solution of two-magnon problem~\cite{SupplemetaryMaterials} is shown in Fig.~3. One can clearly see that with including $d\neq 0$, the bound state minimum of the two-magnon state shifts  from $K=\pi$ to $K=\pi-K_0$, where $K_0\sim d$. The binding energy decreases as well gradually and after the critical value of $d>d_c(J_1)$ (e.g. for $J_1=-1$, $d_c\simeq 0.183$) the two-magnon scattering state minimum becomes energetically lower. Hence, bound states disappear from the ground state.

\begin{figure}[h]
\label{Fig:Twomagnons}
\includegraphics*[width=0.603\textwidth]{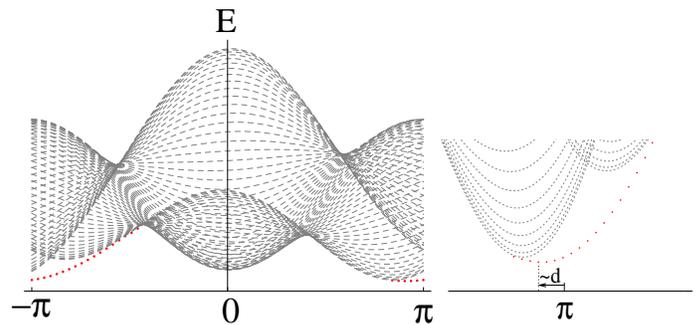}
\caption{ Two particle spectrum, with scattering states and bound state branch for $d=0.15$ and $J_1=-1$. Parity asymmetry is due to DM interaction. Inset shows a zoom of the two-body dispersion around the momentum $\pi$ indicating a shift of the minimum from $\pi$ in the direction of the two-magnon scattering state minimum.}
\end{figure}

When the density of magnons is increased with decreasing the magnetic field we expect that the two-magnon bound states quasi-condense in the minimum of the two-magnon dispersion at $K=\pi-K_0$. Hence, the ground state will enter the nematic-chiral state for an arbitrary small $d\neq 0$. However, when $d>d_c$, the nematicity (magnon pair quasi-condensate) disappears via QPT, and the low energy behavior is dominated by a single-particle picture with $\langle S_i^- S_j^+\rangle $ quasi long-range ordered as shown in right panel of Fig. 4. Hence, we anticipate an $E$-field driven phase transition    from  the 'molecular' (2-magnon bound state) quasi-condensate to  the 'atomic' (single-particle) quasi-condensate. This expectation is fully confirmed by the effective field theory description within bosonization techniques\cite{SupplemetaryMaterials} where the competition between ferromagnetic $J_1$ (that binds magnons and produces nematic order) and electric field (promoting chirality) is resolved via an Ising QPT with changing $d$.

We have checked our analytical results with large scale numerical calculations using the density matrix renormalization group (DMRG) method \cite{white92b,schollwoeck05} on chains up to $L=240$ sites.

\begin{figure}[h]\label{Fig.15}
\includegraphics*[width=0.355\textwidth]{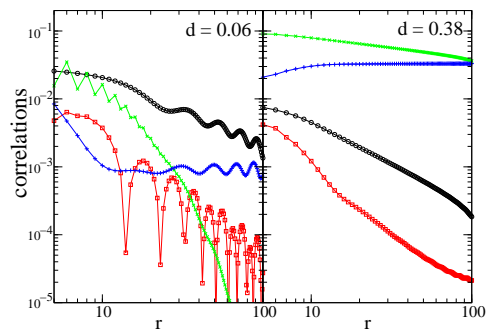}
\caption{(Color online)
Various correlation functions for $J_1=-1$ and $M=0.4$ in nematic (left) and chiral (right) phases for $L=160$ sites. In-plane spin-spin correlation functions $\left<S^+_i S^-_{i+r}\right>$ are indicated by $\times$ and show exponential decay in nematic phase and algebraic quasi-long-range order in the chiral phase; $+$ indicates chirality correlations $\left<\kappa_i\kappa_{i+r}\right>$, pair correlation $\left<S^+_i S^+_{i+1} S^-_{i+r} S^-_{i+r+1}\right>$ indicated by  $\circ$ and the density correlations $\left<S^z_i S^z_{i+r}\right>$ indicated by open squares decay algebraically in both phases with pronounced oscillations in nematic phase.}
\end{figure}

\begin{figure}[h]\label{Fig.12}
\includegraphics*[width=0.45\textwidth]{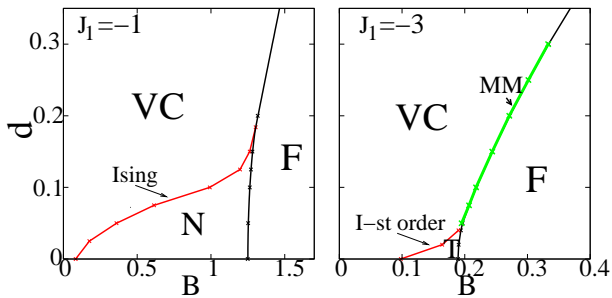}
\caption{(Color online) Phase diagram as a function of electric and magnetic fields. MM indicates metamagnetic behavior (macroscopic jump) in the magnetization when descending from a saturation value. T indicates multipolar state with three-body bound states. We determined phase boundary between nematic (N) and chiral (VC) states by looking at magnetization step size with $B$ for finite systems. $\Delta M=2$ in N, whereas $\Delta M=1$ in VC.
Similarly, we observe phase boundary between T (with magnetization step $\Delta M=3$) and a VC.}
\end{figure}
\begin{figure}[h]\label{Fig.14}
\includegraphics*[width=0.235\textwidth]{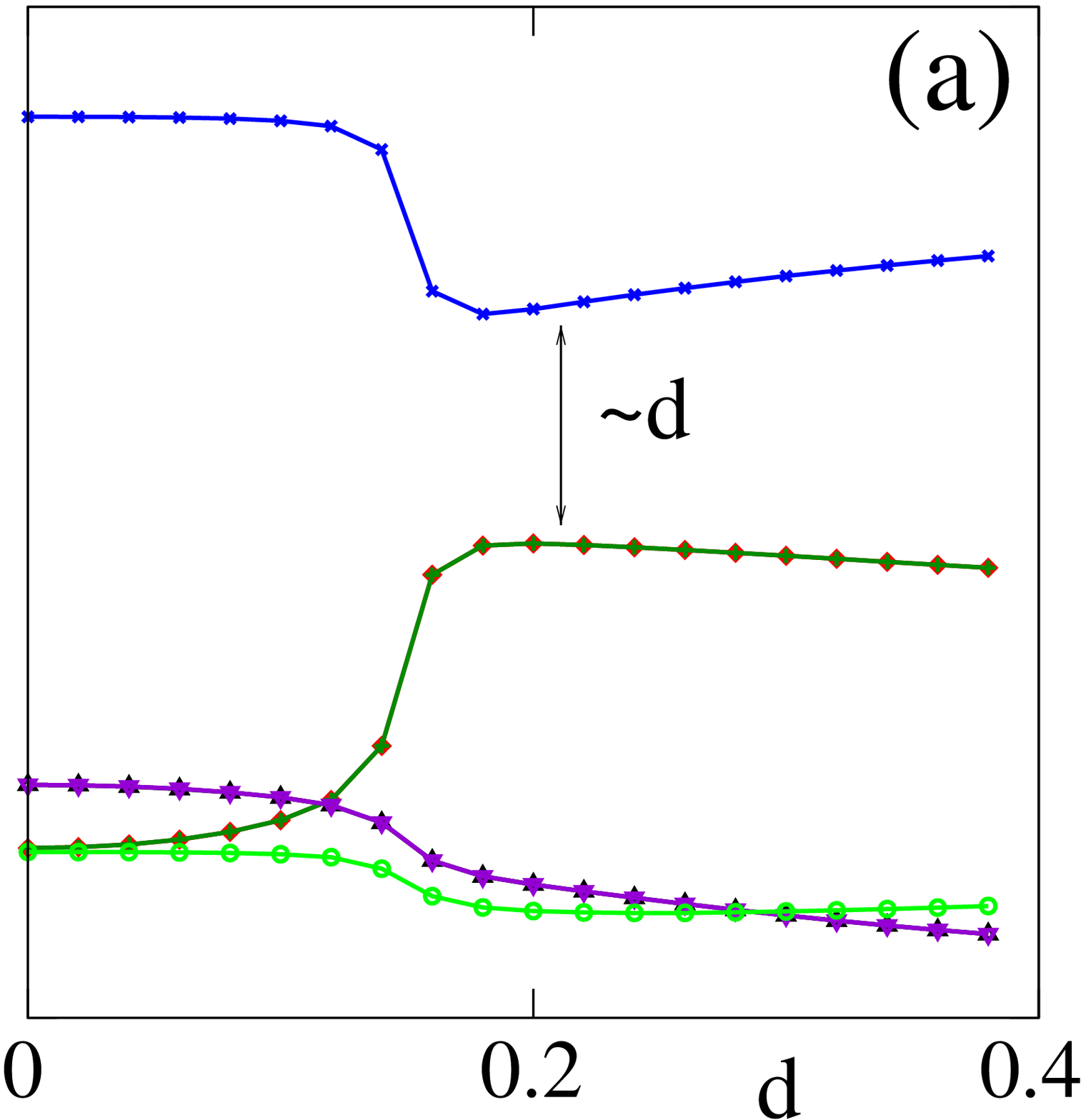}\includegraphics*[width=0.25\textwidth]{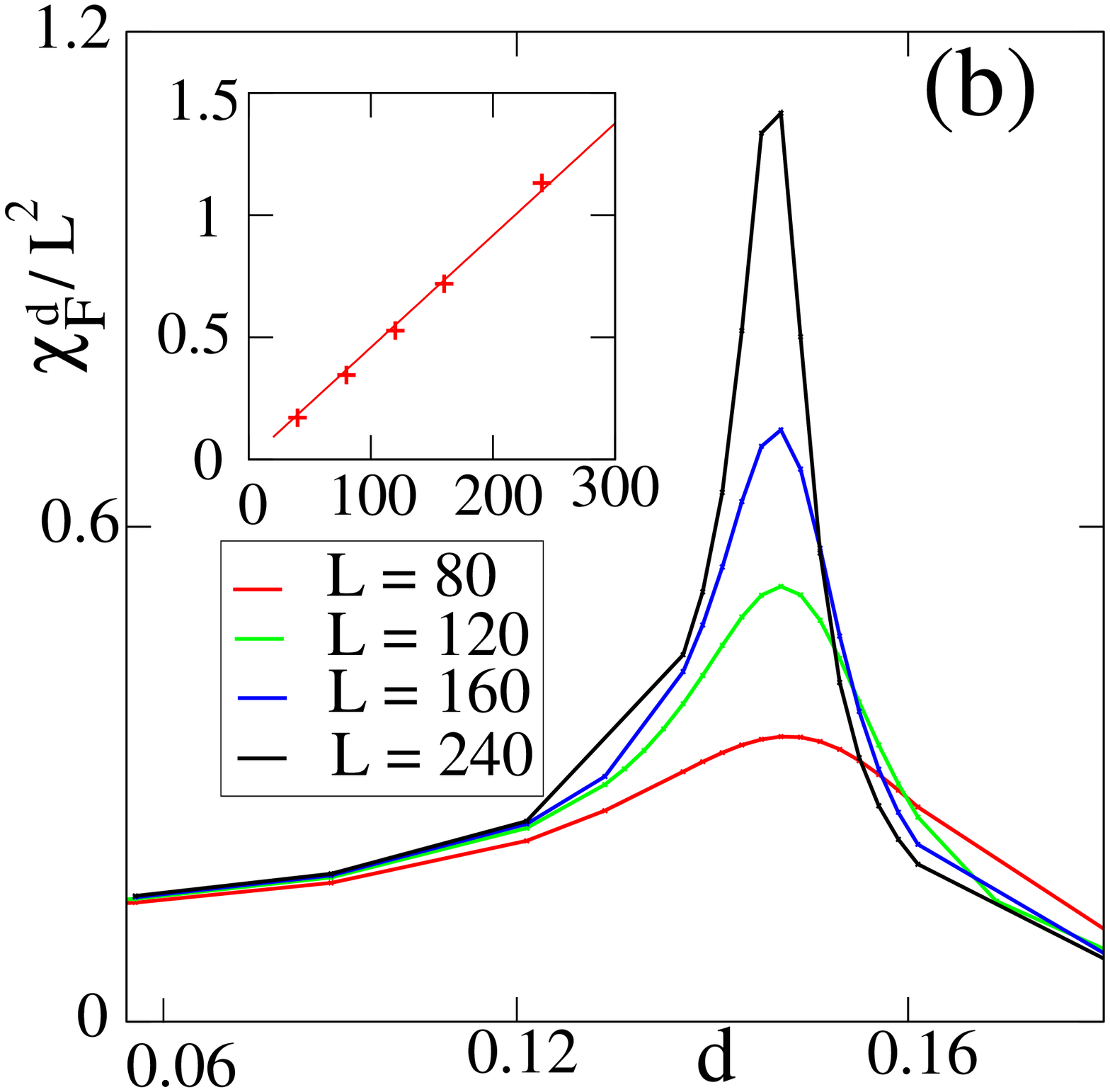}
\caption{(Color online) a) Entanglement spectrum for L=160 sites b) Scaling of DM FS near the nematic to chiral QPT for $J_1=-1$ and $M=0.4$.}
\end{figure}
\begin{figure}[h]\label{Fig.13}
\includegraphics*[width=0.35\textwidth]{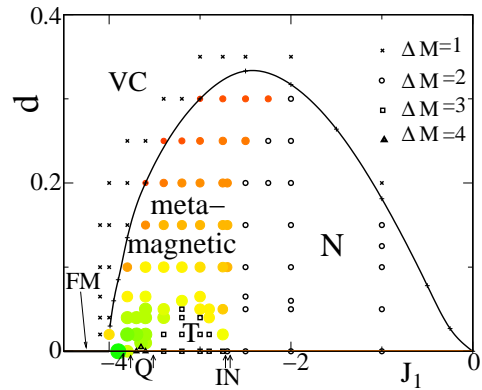}
\caption{ (Color online) Phases under saturation magnetization. N, T and Q stand for multipolar phases with two, three and four-magnon bound states, respectively and IN stands for incommensurate nematic phase. Filled circles indicate that for these parameters the system experiences macroscopic magnetization jump when descending from a fully polarized ground state into VC state by lowering $B$ (indicated by MM in Fig. 5), larger circle meaning greater jump.}
\end{figure}

For small values of $|J_1|\lesssim 2$ and for $d=0$ the leading correlation function is $\langle S_i^z S_j^z\rangle $ for low magnetic fields $B\neq 0$ and the system is in the SDW dominated regime. With increasing the magnetic field SDW phase crosses over into the nematic state\cite{vekua07}, with leading correlation function given by $\langle S_i^-S_{i+1}^-\, S_j^+ S_{j+1}^+\rangle $ (see left pannel of Fig. 4). In both regimes the in-plane single-spin correlation function $\langle S_i^- S_j^+\rangle $ decays exponentially. We have studied various correlation functions for different values of electric field. In Fig. 4 we compare the behavior of the correlation functions in nematic ($d<d_c$) and chiral ($d>d_c$) phases.
In Fig. 5 we depict the phase diagram as a function of driving fields $E$ and $B$ at $J_1=-1$ (a) and $J_1=-3$ (b).
To witness the transition from the nematic to the chiral state induced by $E$ we studied the behavior of the entanglement spectrum (Fig. 6 (a)) and DM FS (Fig. 6 (b)). In the chiral phase of a $J_1-J_2$ chain and for $d=0$ the complete entanglement spectrum is doubly degenerate due to the spontaneously broken parity symmetry, however in the presence of $d$  the degeneracy is lifted. Linear in $L$ scaling of the peak of DM FS relative to the overall background shown in inset of Fig. 6 b) confirms the Ising nature of  QPT.

We have studied as well the effect of DM anisotropy on multipolar phases of the $J_1-J_2$ chain for $-4<J_1<-2.7$ involving bound states with more than 2 magnons. The minimum of the multi-body bound state dispersion which is at $K=\pi$ for $d=0$ (in both phases T and Q) shifts from $\pi$  for $d\neq 0$. In fact, $1 \% \sim 2\%$ DM anisotropy in $J_1$ is sufficient to remove the three-body and the four-body multipolar phases from the ground state phase diagram below the saturation magnetization. Instead, in the presence of a tiny $d\neq 0$ the ground state magnetization experiences a macroscopic jump to the fully saturated value when increasing the magnetic field as depicted in Fig. 7. Note, for $d=0$ the metamagnetic region is squeezed in the close right-side vicinity of $J_1=-4$ point. In the presence of DM anisotropy the metamagnetic jump is observed in much broader region, starting at $J_1\simeq -2$ and extending even in the region $J_1<-4$ \cite{SupplemetaryMaterials}.

{\it In summary},  based on the spin current model for a helical multiferroic  spin-$\frac{1}{2}$ chain in external $B$ and $E$-fields
  we find that both quantum information processing as well as ground state phases are extremely sensitive to an electric field that affects the magnetoelectric coupling.
$E$-field  increases strongly the quantum state transfer fidelity and makes it directional dependent (transfer in clockwise direction differs from that in anticlockwise direction). A tiny   magnetoelectric coupling  is sufficient to change the spin-density-wave/nematic or multipolar phases in favor of the chiral phase.
We analyzed QPT induced by ME coupling and find in particular a sharp change of the entanglement spectrum and a large enhancement of the fidelity susceptibility at Ising QPT from nematic to chiral states. Our findings serve as the basis for $E$ field controlled   quantum information processing in helical multiferoics.

\acknowledgements
MA, LC, SKM and JB acknowledge gratefully  financial support by the Deutsche Forschungsgemeinschaft (DFG) through SFB 762, and contract BE 2161/5-1. SG and TV are supported by QUEST (Center for Quantum Engineering and Space-Time Research) and DFG
Research Training Group (Graduiertenkolleg) 1729.

\section{Details of bosonization}
Here we provide details of effective field theory description, bosonization applied to microscopic Hamiltonian
\begin{eqnarray}
\label{eq.1}
\hat{H}&=&J_1\displaystyle\sum_{i=1}\vec{S}_i.\vec{S}_{i+1}
+J_2\displaystyle\sum_{i=1}\vec{S}_i.\vec{S}_{i+2}\nonumber\\
&-&B\displaystyle\sum_{i=1}S_{i}^{z}- d \sum_{i=1}(\vec{S}_i\times \vec{S}_{i+1})^z.
\end{eqnarray}

To develop bosonization description it is convenient to consider the limit of strong frustration $J_2 \gg |J_1|$ and weak DM anisotropy $d\ll J_2 $. In this case
the system may be viewed as two antiferromagnetic spin-$\frac{1}{2}$ chains weakly coupled by the zigzag interchain coupling $J_{1}$ \cite{Nersesyan} with DM anisotropy $d$.

Low-energy properties of a single spin-$\frac{1}{2}$ chain in a uniform magnetic field is described by the standard
Gaussian theory \cite{LutherPeschel} known also as the Tomonaga-Luttinger liquid:
\begin{equation}
\label{SpinChainBosHam}
{\cal H} =  \frac{v}{2}\int dx \, \Big\{\frac{1}{K}(\partial_x \phi)^{2}
+ K (\partial_x \theta)^{2}\Big\}.
\end{equation}
Here $\phi$ is a real scalar bosonic field and $\theta$ is its
dual field, $\partial_t \phi =v \partial_x \theta $, with the commutation
relations $[\phi(x),\theta(y)] = i\Theta (y-x)$, where $\Theta(x)$ is the
Heaviside function. $K$ is Luttinger liquid parameter and $v$ is spin-wave velocity.

 The exact functional dependences $v(J_2,B)$
and $K(J_2,B)$ for isolated chains are known (see \cite{AffleckOshikawa} and references therein) from the numerical solution   of
the Bethe ansatz integral equations \cite{Bogoliubov}. In particular, $K$ increases monotonously
with the magnetic field, whereas $v$ decreases: $K(B=0)=\frac{1}{2}$, $v(B=0)=J_2\pi/2$ and $K\to 1$, $v\to 0$ for $B\to B_{sat}$, where saturation value $B_{sat}=2J_2$.

Long wave-length fluctuations of spin-1/2 chain are captured by
the following representation of the lattice spin operators \cite{LutherPeschel}:
\begin{eqnarray}
\label{Luther}
S^z_n&\to &\frac{1}{\sqrt{\pi}}\partial_x \phi
+\frac{a}{\pi} \sin \big\{2k_Fx+ \sqrt{4\pi } \phi\big\}  +M\\
S_n^-& \to & (-1)^n e^{-i\theta\sqrt{\pi}}
\big\{ c+b\sin{\big(2k_F x+ \sqrt{4\pi }\phi\big)}\big\},\nonumber
\end{eqnarray}
Here $M(B)$ is the ground state magnetization per
spin which determines the Fermi wave vector $k_F=(\frac{1}{2}-M)\pi$ and $a$, $b$, and $c$ are non-universal numerical constants.

For $J_1=d=0$, two decoupled
chains are described by two copies of
Gaussian models of the form (\ref{SpinChainBosHam}) with pair of dual bosonic fields $[\phi_1,\theta_1]$ and $[ \phi_2, \theta_2]$ .
 Treating interchain couplings $J_1$ and DM anisotropy $d$ perturbatively and introducing the symmetric and
antisymmetric combinations of the fields describing the individual chains,
$\phi_{\pm}=(\phi_1\pm \phi_2)/\sqrt{2K}$ and
$\theta_{\pm}= (\theta_1 \pm \theta_2)\sqrt{K/2}$,
the effective Hamiltonian density describing low-energy properties of (\ref{eq.1}) takes the following form:
\begin{eqnarray}
\label{symantisym}
{\mathcal H}_{\rm eff}  &=& {\mathcal H}_{0}^{+} +{\mathcal H}_{0}^{-} +
{\mathcal H}_{\rm int},\nonumber\\
{\mathcal H}_{0}^{\pm} & =&\frac{v_{\pm}}{2} [(\partial_x \theta_{\pm})^{2} + (\partial_x
  \phi_{\pm})^2], \nonumber\\
{\mathcal H}_{\rm int} &=& g_1 \cos\big(k_F+\sqrt{8\pi K_-}\phi_{-}\big)
\nonumber\\
&-& (g_2\partial_x \theta_+ +g_3)\sin\big(\sqrt{2\pi/K_{-}}\theta_-\big).
\end{eqnarray}

The Fermi velocities $v_{\pm} \propto J_{2}$ and coupling constants are
$g_{1}\propto J_1\cos{k_F}$ \cite{Cabra}, $g_{2}\propto J_1$ and $g_3\propto d$, with proportionality coefficients involving short-distance cut-off. The Luttinger Liquid parameter of antisymmetric sector is given by
\begin{equation}
K_-=K(h)\Big\{ 1 + J_1 K(B)/\big(\pi v(B)\big) \Big\}.
\end{equation}

The inter-sector coupling in Eq.~(\ref{symantisym}) contains a term with coupling constant $g_2$ that represents an infrared limit of the
product of z-components of in-chain and inter-chain vector chiralities \cite{Lecheminant},
\begin{equation}
\label{chirprod}
 (\kappa^{z}_{2i-1,2i+1}+ \kappa^{z}_{2i,2i+2})\,\kappa^{z}_{2i,2i+1} \to \partial_x \theta_+\sin\sqrt {\frac{2\pi}{ K_-}}\theta_- ,
\end{equation}
where $\kappa_{i,j}^{z}\equiv (\vec{S}_{i}\times \vec{S}_{j})^{z}$.

 The Hamiltonian
(\ref{symantisym}) provides with the effective field theory
 describing the low-energy behavior of a strongly
frustrated spin-$\frac{1}{2}$ zigzag chain with DM anisotropy for a nonzero
magnetization $M$. For small values of magnetization the Luttinger liquid parameter $K_{-}\simeq \frac{1}{2}$, and
the inter-sector $g_{2}$ term has a higher scaling dimension than the strongly
relevant $g_{1}$ and $g_3$ terms in the antisymmetric
sector. In this case the system is in a phase with relevant competing couplings in
antisymmetric sector. In contrast to that, at $B=0$ all terms generated
by the $J_1$ zigzag coupling are marginal and only DM coupling $g_3$ is a relevant perturbation.

The competition between $\cos {\sqrt{8\pi K_- }\phi_-}$ (nematicity) and $\cos\sqrt{2\pi/K_-} \theta_-$ (chirality) terms is resolved with an Ising phase transition in the antisymmetric sector with changing $d/J_1$ \cite{Gogolin}.

\section{Effect of Dm anisotropy in ferromagnetic region $J_1<-4J_2$}

We now discuss the effect of DM interaction on ferromagnetic region $J_1<-4J_2$. For $d=0$, due to SU(2) symmetry the magnon gas behaves as non-interacting bosons. Deep inside ferromagnetic region DM interaction introduces repulsion (repulsion increases monotonously with increasing $d$) between magnons and below the fully polarized state chiral Luttinger liquid phase is realized for any $d \neq 0$ \cite{Mahdavifar}. However, in close left-side vicinity of $J_1 = -4$ (hence $J_1<-4$) non-monotonous effect of DM on the effective interaction between magnons is observed. First, for small values $d\to 0$ DM anisotropy introduces repulsion between magnons, however with increasing $d$ repulsion transforms into attraction and with further increasing $d$ interaction between magnons becomes repulsive once again as shown in Fig. 1. Effective coupling constant of the magnon gas we extracted from the following relation \cite{Vekua1,Vekua2},
\begin{equation}
\label{efint}
g =-\frac{2 \hbar^2}{m a_{1D}}
\end{equation}

where $m$ is mass of magnon and $a_{1D}$ is one-dimensional scattering length, which we calculated analytically from the low energy scattering phase shift $\delta(k)$,
\begin{equation}
a_{1D}=\lim_{k\to 0} \frac{\delta(k)}{k},
\end{equation}
where $k$ is a relative momentum of scattering magnons.

\begin{figure}[h]\label{Fig.10}
\includegraphics*[width=0.45\textwidth]{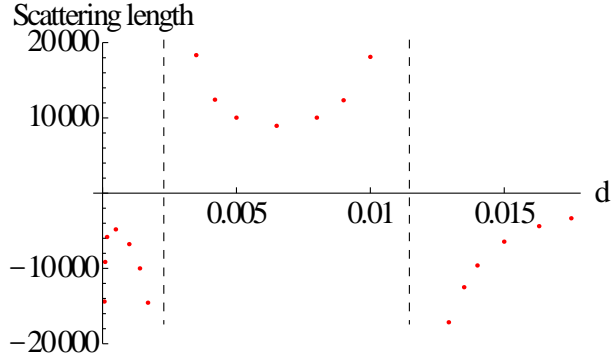}
\caption{Two-magnon scattering length in units of the lattice constant for $J_1=-4.005$, showing non-trivial sequence of resonances induced by changing just a single parameter $d$.}
\end{figure}

For attractive regime $g<0$, $a_{1D}>0$, scattering length extracted from scattering problem coincides with the correlation length of the bound state of magnons.
We depict in Fig. 1 scattering length from which one can observe due to Eq. (\ref{efint}) that effective interaction changes sign twice via resonance-like behavior when changing $d$. For the values of $d$ which correspond to the positive scattering length (and hence $g<0$), the external magnetic field induces a metamagnetic transition (macroscopic jump of the magnetization) from chiral Luttinger liquid to the fully polarized state (resulting in first order phase transition). For the parameters corresponding to negative scattering length (and hence $g>0$) magnetization will change smoothly all the way from $M=0$ till $M=1/2$, in particular leading to usual commensurate-incommensurate phase transition from chiral Luttinger liquid to fully polarized state when increasing the magnetic field strength.

\section{Fidelity}

Transition amplitudes and energy levels entering in the expression for fidelity Eq. (2), used for plotting  Fig. 1:

\begin{eqnarray}\label{eq.11}
f_{1,2}=\frac{1}{4}(\exp[-i\wp_5t]-\exp[-i\wp_4t]) \nonumber
\\-\frac{i}{4}(\exp[-i\wp_2t]-\exp[-i\wp_3t]),\nonumber
\end{eqnarray}

\begin{eqnarray}\label{eq.12}
f_{2,1}=-\frac{1}{4}(\exp[-i\wp_4t]-\exp[-i\wp_5t]) \nonumber
\\-\frac{i}{4}(\exp[-i\wp_2t]-\exp[-i\wp_3t]),
\end{eqnarray}

\begin{eqnarray}\label{eq.13}
f_{1,3}=f_{3,1}=-\frac{1}{4}(\exp[-i\wp_2t]+\exp[-i\wp_3t] \nonumber
\\-\exp[-i\wp_4t]-\exp[-i\wp_5t]),\nonumber
\end{eqnarray}

\begin{eqnarray}\label{eq.14}
\wp_{2}=-J_2-B-d,\wp_{3}=-J_2-B+d,\nonumber\\
\wp_{4}=-J_1+J_2-B,\wp_{5}=J_1+J_2-B.
\end{eqnarray}

\end{document}